\documentclass{IEEEtran}
\usepackage{amsmath,amsfonts}
\usepackage{algorithmic}
\usepackage{algorithm}
\usepackage{array}
\usepackage[caption=false,font=normalsize,labelfont=sf,textfont=sf]{subfig}
\usepackage{textcomp}
\usepackage{stfloats}
\usepackage{verbatim}
\usepackage{graphicx}
\usepackage{cite}
\usepackage{tabularx}
\usepackage{booktabs}
\usepackage{acronym}
\acrodef{RIS} [RIS] {reconfigurable intelligent surface}
\acrodef{gnss} [GNSS] {global navigation satellite systems}
\acrodef{star} [STAR] {simultaneously transmitting and reflecting}
\acrodef{CRLB} [CRLB] {Crámer-Rao lower bound}
\acrodef{BF} [BF] {beamforming}
\acrodef{SNR} [SNR] {signal-to-noise ratio}
\acrodef{los} [LoS] {line-of-sight}
\acrodef{nlos} [NLoS] {non-line-of-sight}
\acrodef{AoD} [AoD] {angle-of-departure}
\acrodef{CFO} [CFO] {carrier frequency offset}
\acrodef{peb} [PEB] {positioning error bound}
\acrodef{ue} [UE] {user equipment}
\acrodef{aod} [AoD] {angle-of-departure}
\acrodef{aoa} [AoA] {angle-of-arrival}
\acrodef{3d} [3D] {three-dimensional}
\acrodef{9d} [9D] {nine-dimensional}
\acrodef{ukf} [UKF] {unscented Kalman filter}
\acrodef{5g} [5G] {fifth-generation}
\acrodef{6g} [6G] {sixth-generation}
\acrodef{soop} [SOOP] {signals of opportunity}
\acrodef{dmimo}[D-MIMO]{distributed MIMO}
\acrodef{MIMO}{multiple-input multiple-output}
\acrodef{bs}[BS]{base station}
\acrodef{ap}[AP]{access point}
\acrodef{NTN}{non-terrestrial network}
\acrodef{TN}{terrestrial network}
\acrodef{uav}[UAV]{unmanned aerial vehicle}
\acrodef{LEO}{low Earth orbit}
\acrodef{MEO}{medium Earth orbit}
\acrodef{GEO}{geosynchronous Earth orbit}
\acrodef{HAPS}{high-altitude platform station}
\acrodef{GDoP}{geometric dilution of precision}
\acrodef{DL}{downlink}
\acrodef{UL}{uplink}
\acrodef{PRS}{positioning reference signals}
\acrodef{SRS}{sounding reference signals}
\acrodef{TDOA}{time difference of arrival}
\acrodef{RTOA}{relative time of arrival}
\acrodef{e-CID}{enhanced cell-ID}
\acrodef{IoT}{internet of things}
\acrodef{RTT}{round-trip time}
\acrodef{TRP}{transmission reception point}
\acrodef{LMF}{location management function}
\acrodef{PNT}{positioning, navigation, and timing}
\acrodef{SoOP}{signals of opportunity}
\acrodef{CFO}{carrier frequency offset}
\acrodef{OFDM}{orthogonal frequency division multiplexing}
\usepackage{color}

\usepackage{tabu,longtable}
\usepackage{diagbox}
\usepackage{threeparttable}
\usepackage{makecell}
\usepackage[table]{xcolor}
\usepackage{multirow} % enable table multirow
\usepackage{tikz} 
\usepackage[utf8]{inputenc}
\usepackage{pgfplots}
\pgfplotsset{compat=1.18} %requested by overleaf? {Sharief}
\usepackage[bottom=1.04in,top=0.75in,left=0.625in,right=0.625in]{geometry}
\usepackage{units} 		% use \unit command
\usepackage{bm} 		% for bold symbol
\usepackage{amssymb} 	% for AMS symbols

\begin{document}
% transpose and hermitian
\newcommand{\TT}{\mathsf{T}}
\newcommand{\HH}{\mathsf{H}}

% Vectors
\newcommand{\av}{{\bf a}}
\newcommand{\bv}{{\bf b}}
\newcommand{\cv}{{\bf c}}
\newcommand{\dv}{{\bf d}}
\newcommand{\ev}{{\bf e}}
\newcommand{\fv}{{\bf f}}
\newcommand{\gv}{{\bf g}}
\newcommand{\hv}{{\bf h}}
\newcommand{\iv}{{\bf i}}
\newcommand{\jv}{{\bf j}}
\newcommand{\kv}{{\bf k}}
\newcommand{\lv}{{\bf l}}
\newcommand{\mv}{{\bf m}}
\newcommand{\nv}{{\bf n}}
\newcommand{\ov}{{\bf o}}
\newcommand{\pv}{{\bf p}}
\newcommand{\qv}{{\bf q}}
\newcommand{\rv}{{\bf r}}
\newcommand{\sv}{{\bf s}}
\newcommand{\tv}{{\bf t}}
\newcommand{\uv}{{\bf u}}
\newcommand{\wv}{{\bf w}}
\newcommand{\vv}{{\bf v}}
\newcommand{\xv}{{\bf x}}
\newcommand{\yv}{{\bf y}}
\newcommand{\zv}{{\bf z}}
\newcommand{\zerov}{{\bf 0}}
\newcommand{\onev}{{\bf 1}}
\newcommand{\avr}{\av_\text{R}}
% Matrices
\newcommand{\Am}{{\bf A}}
\newcommand{\Bm}{{\bf B}}
\newcommand{\Cm}{{\bf C}}
\newcommand{\Dm}{{\bf D}}
\newcommand{\Em}{{\bf E}}
\newcommand{\Fm}{{\bf F}}
\newcommand{\Gm}{{\bf G}}
\newcommand{\Hm}{{\bf H}}
\newcommand{\Id}{{\bf I}}
\newcommand{\Jm}{{\bf J}}
\newcommand{\Km}{{\bf K}}
\newcommand{\Lm}{{\bf L}}
\newcommand{\Mm}{{\bf M}}
\newcommand{\Nm}{{\bf N}}
\newcommand{\Om}{{\bf O}}
\newcommand{\Pm}{{\bf P}}
\newcommand{\Qm}{{\bf Q}}
\newcommand{\Rm}{{\bf R}}
\newcommand{\Sm}{{\bf S}}
\newcommand{\Tm}{{\bf T}}
\newcommand{\Um}{{\bf U}}
\newcommand{\Wm}{{\bf W}}
\newcommand{\Vm}{{\bf V}}
\newcommand{\Xm}{{\bf X}}
\newcommand{\Ym}{{\bf Y}}
\newcommand{\Zm}{{\bf Z}}
\newcommand{\Onem}{{\bf 1}}
\newcommand{\Zerom}{{\bf 0}}
% text uppercase
\newcommand{\At}{{\rm A}}
\newcommand{\Bt}{{\rm B}}
\newcommand{\Ct}{{\rm C}}
\newcommand{\Dt}{{\rm D}}
\newcommand{\Et}{{\rm E}}
\newcommand{\Ft}{{\rm F}}
\newcommand{\Gt}{{\rm G}}
\newcommand{\Ht}{{\rm H}}
\newcommand{\It}{{\rm I}}
\newcommand{\Jt}{{\rm J}}
\newcommand{\Kt}{{\rm K}}
\newcommand{\Lt}{{\rm L}}
\newcommand{\Mt}{{\rm M}}
\newcommand{\Nt}{{\rm N}}
\newcommand{\Ot}{{\rm O}}
\newcommand{\Pt}{{\rm P}}
\newcommand{\Qt}{{\rm Q}}
\newcommand{\Rt}{{\rm R}}
\newcommand{\St}{{\rm S}}
\newcommand{\Tt}{{\rm T}}
\newcommand{\Ut}{{\rm U}}
\newcommand{\Wt}{{\rm W}}
\newcommand{\Vt}{{\rm V}}
\newcommand{\Xt}{{\rm X}}
\newcommand{\Yt}{{\rm Y}}
\newcommand{\Zt}{{\rm Z}}

% Bold greek letters
\newcommand{\alphav}{\hbox{\boldmath$\alpha$}}
\newcommand{\betav}{\hbox{\boldmath$\beta$}}
\newcommand{\gammav}{\hbox{\boldmath$\gamma$}}
\newcommand{\deltav}{\hbox{\boldmath$\delta$}}
\newcommand{\etav}{\hbox{\boldmath$\eta$}}
\newcommand{\lambdav}{\hbox{\boldmath$\lambda$}}
\newcommand{\kappav}{\hbox{\boldmath$\kappa$}}
\newcommand{\epsilonv}{\hbox{\boldmath$\epsilon$}}
\newcommand{\nuv}{\hbox{\boldmath$\nu$}}
\newcommand{\muv}{\hbox{\boldmath$\mu$}}
\newcommand{\zetav}{\hbox{\boldmath$\zeta$}}
\newcommand{\phiv}{\hbox{\boldmath$\phi$}}
\newcommand{\psiv}{\hbox{\boldmath$\psi$}}
\newcommand{\thetav}{\hbox{$\boldsymbol\theta$}}
\newcommand{\tauv}{\hbox{\boldmath$\tau$}}
\newcommand{\omegav}{\hbox{\boldmath$\omega$}}
\newcommand{\xiv}{\hbox{\boldmath$\xi$}}
\newcommand{\sigmav}{\hbox{\boldmath$\sigma$}}
\newcommand{\piv}{\hbox{\boldmath$\pi$}}
\newcommand{\rhov}{\hbox{\boldmath$\rho$}}

\newcommand{\Gammam}{\hbox{\boldmath$\Gamma$}}
\newcommand{\Lambdam}{\hbox{\boldmath$\Lambda$}}
\newcommand{\Deltam}{\hbox{\boldmath$\Delta$}}
\newcommand{\Sigmam}{\hbox{\boldmath$\Sigma$}}
\newcommand{\Phim}{\hbox{\boldmath$\Phi$}}
\newcommand{\Pim}{\hbox{\boldmath$\Pi$}}
\newcommand{\Psim}{\hbox{\boldmath$\Psi$}}
\newcommand{\psim}{\hbox{\boldmath$\psi$}}
\newcommand{\chim}{\hbox{\boldmath$\chi$}}
\newcommand{\omegam}{\hbox{\boldmath$\omega$}}
\newcommand{\Thetam}{\hbox{\boldmath$\Theta$}}
\newcommand{\Omegam}{\hbox{\boldmath$\Omega$}}
\newcommand{\Xim}{\hbox{\boldmath$\Xi$}}

\bstctlcite{IEEEexample:BSTcontrol}

% \title{6G Unified Localization: \\The Integration of TN and NTN}
\title{Integrated 6G TN and NTN Localization: \\Challenges, Opportunities, and Advancements}

\author{Sharief Saleh,~\IEEEmembership{Member,~IEEE},
Pinjun~Zheng,~\IEEEmembership{Graduate Student Member,~IEEE},
Xing~Liu,~\IEEEmembership{Member,~IEEE},\\
Hui~Chen,~\IEEEmembership{Member,~IEEE},
Musa~Furkan~Keskin,~\IEEEmembership{Member,~IEEE},
Basuki~Priyanto,~\IEEEmembership{Senior Member,~IEEE},\\
Martin~Beale,
Yasaman~Ettefagh,~\IEEEmembership{Graduate Student Member,~IEEE},
Gonzalo~Seco-Granados,~\IEEEmembership{Fellow,~IEEE},\\
Tareq~Y.~Al-Naffouri,~\IEEEmembership{Fellow,~IEEE},
and~Henk Wymeersch,~\IEEEmembership{Fellow,~IEEE}\vspace{-2em}

\thanks{S. Saleh, H. Chen, F. Keskin, Y. Ettefagh, and H. Wymeersch are with the Department of Electrical Engineering, Chalmers University of Technology, 412 58 Gothenburg, Sweden (Email: \{sharief; hui.chen; furkan; ettefagh; henkw\}@chalmers.se).  
P.~Zheng, X.~Liu, and T.~Y.~Al-Naffouri are with the Electrical and Computer Engineering Program, Division of Computer, Electrical and Mathematical Sciences and Engineering (CEMSE), King Abdullah University of Science and Technology (KAUST), Thuwal, 23955-6900, Kingdom of Saudi Arabia (Email: \{pinjun.zheng; xing.liu; tareq.alnaffouri\}@kaust.edu.sa). B. Priyanto and M. Beale are with Sony Europe BV (Email: \{basuki.priyanto; martin.beale\}@sony.com). 
G. Seco-Granados is with the Department of Telecommunications and Systems Engineering, Universitat Autònoma de Barcelona, 08193 Bellaterra, Spain (E-mail: gonzalo.seco@uab.cat).}

\thanks{This work is supported by the European Commission through the Horizon Europe/JU SNS project Hexa-X-II (Grant Agreement no. 101095759), the Swedish Research Council through Grants 2022-03007 and 2024-04390, the Spanish R+D project PID2023-152820OB-I00, the Catalan ICREA Academia Program, the King Abdullah University of Science and Technology (KAUST) Office of Sponsored Research (OSR) under Award ORA-CRG2021-4695, and Vinnova FFI project 2023-02603.}
}
%\markboth{This work has been submitted to the IEEE for possible publication.  Copyright may be transferred without notice.}{draft}
\maketitle
\thispagestyle{empty}
\pagestyle{empty}

\begin{abstract}
The rapid evolution of cellular networks has introduced groundbreaking technologies, including large and distributed antenna arrays and reconfigurable intelligent surfaces in terrestrial networks (TNs), as well as aerial and space-based nodes in non-terrestrial networks (NTNs). These advancements enable applications beyond traditional communication, such as high-precision localization and sensing. While integrating TN and NTN enablers will lead to unparalleled opportunities for seamless global localization, such integration attempts are expected to face several challenges. To understand these opportunities and challenges, we first examine the distinctive characteristics of the key 6G enablers, evaluating their roles in localization from both technical and practical perspectives. Next, to identify developments driving TN-NTN localization, we review the latest standardization and industrial innovation progress. Finally, we discuss the opportunities and challenges of TN-NTN integration, illustrating its potential through two numerical case studies. % that demonstrate its impact on global positioning systems.
\end{abstract}


%%%%%%%%%%%%%%%%%%%%%%%%%%%%%%%%%%%%%%%%%%%%%%%%%%%%%%%%%%%%%%%
\section{Introduction}
% The rapid evolution of wireless communication technologies has brought us to a new era with the anticipated arrival of 6G networks. 
With the anticipated arrival of 6G, we stand on the brink of a transformative generation in advanced communication services. Like its predecessors, 6G aims to deliver higher data rates, ultra-low latency, ubiquitous connectivity, and enhanced security~\cite{dang2020should}. The foundational enablers of 6G include both \acp{TN} and \acp{NTN}. While \acp{TN} serve as the foundation of the global communication infrastructure, \acp{NTN} are set to complement \acp{TN} and extend connectivity into remote and rural areas. Thus, the integration of \acp{TN} and \acp{NTN} will create a robust three-dimensional (3D) network that is essential to achieve the ambitious goals of 6G communication. 
However, communications is not the only goal of future 6G networks. Over the past years, cellular communication systems have increasingly integrated positioning services, a trend that is becoming even more pronounced in 6G~\cite{9976205}. Positioning information is essential for a wide range of applications, including autonomous vehicles, smart cities, industrial \ac{IoT}, emergency response systems, and augmented reality. %, as well as internal 6G services such as \ac{ue}-based sensing and context-aided communication. 
These applications impose rigorous requirements on positioning systems, demanding not only high accuracy, typically ranging from meters to centimeters, but also wide coverage, high reliability, low latency, and resilience in challenging environments. Generally, existing \acp{TN} and \acp{NTN} cannot fully meet these demanding requirements independently. For example, the coverage of \acp{TN} is constrained by the availability and density of infrastructure, whereas satellite signals in \acp{NTN} experience significant degradation due to obstruction and attenuation in dense urban areas~\cite{9617565}. Addressing these challenges requires \textit{an integrated approach that leverages the strengths of both \acp{TN} and \acp{NTN}} to deliver precise and reliable positioning under diverse conditions.

To understand \ac{TN}-\ac{NTN} integration from technical and standardization perspectives, we begin by analyzing the characteristics of key 6G enablers in both networks and their interactions across various 6G frequency ranges (from FR1 to sub-THz frequencies). This analysis will help identify the most suitable combinations of enablers and frequency ranges for effective integration. We then review the latest 3GPP standardization efforts for \ac{TN} and \ac{NTN} positioning, alongside recent advancements in proprietary \ac{NTN} \ac{PNT} solutions, shedding light on anticipated industrial trends. Next, we highlight the challenges and opportunities of integrating \ac{TN} and \ac{NTN}, providing the research community with a list of key problems that require immediate attention. Finally, we present two numerical case studies of tightly integrated \ac{TN}-\ac{NTN} systems, further demonstrating the potential of this integration.

% In this work, we explore the integration of \acp{TN} and \acp{NTN} to enhance localization in 6G networks. We highlight how the synergy of these technologies can improve positioning and tracking performance. However, despite the promising advancements, the integration of 6G \acp{TN} with \acp{NTN} remains in its early stages and presents numerous challenges, including the need for interoperability and standardization across various platforms, the management of positioning latency (which is critical for real-time applications), and the mitigation of various sources of errors and imperfections (e.g., atmospheric effects and synchronization biases). These obstacles must be overcome to harness the full potential of this integration in 6G networks. In this paper, we will provide an overview of the characteristics of 6G enablers in both \acp{TN} and \acp{NTN} through the lens of localization, review the 3GPP standardization and industrial efforts, outline the challenges and opportunities inherent in their integration, and showcase two TN-NTN localization case studies. 

\begin{figure*}
    \centering
    \includegraphics[width=0.9\linewidth]{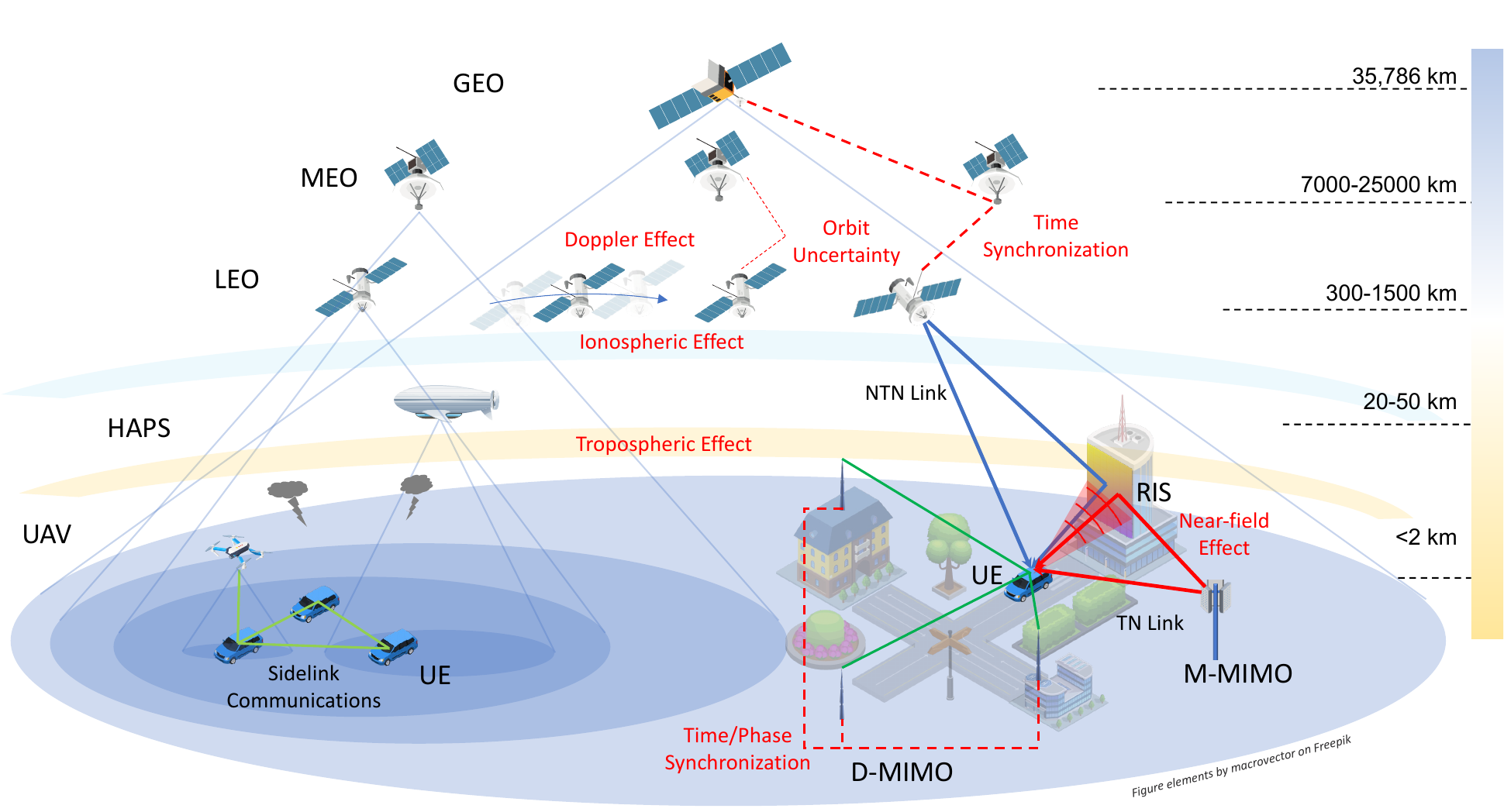}
    \caption{Illustration of various NTN (GEO, MEO, LEO, HAPS, and UAV) and TN enablers (M-MIMO, D-MIMO, RIS, and sideline communications) and their characteristics (Doppler effect, orbit uncertainty, near-field effect, and time/phase synchronization).}
    \label{6G_Enablers}
\end{figure*}

%%%%%%%%%%%%%%%%%%%%%%%%%%%%%%%%%%%%%%%%%%%%%%%%%%%%%%%%%%%%%%%
\section{6G Localization Enablers and Characteristics}
The enablers of 6G localization in both \ac{TN} and \ac{NTN}, shown in Fig.~\ref{6G_Enablers}, come with distinct, yet complementary, characteristics, offering unique opportunities and challenges. These characteristics are shaped by the fundamental properties of these technologies, their operational frequency bands, and the localization environment. For example, the high altitudes of \ac{NTN} systems enable extensive coverage but introduce latency and propagation challenges, distinguishing them from \ac{TN} systems. Similarly, the envisioned operation of 6G across a broad frequency spectrum—from FR1 to FR4\footnote{While FR1 (sub-7.125 GHz) and FR2 (mmWave) are well-defined in the context of cellular communication standards, FR3 (7.125-24.25 GHz) and FR4 (90-300 GHz) are currently being studied and are not yet officially defined by 3GPP.}—further diversifies the behavior and performance of these technologies across various scenarios. This section explores the characteristics of each 6G enabler, emphasizing their interplay with different frequency ranges and environmental conditions.

\subsection{Terrestrial Networks}
Massive \ac{MIMO}, \ac{dmimo}, \acp{RIS}, and sidelink communications play a central role in enabling 6G \ac{TN}-based localization. Each technology involves specific hardware requirements, algorithmic challenges, and benefits, as well as mobility considerations across various frequency bands under unique propagation conditions. The following sections will explore their distinct characteristics.

\subsubsection{Massive MIMO}
M-\ac{MIMO} employs large antenna arrays at the \ac{bs} to simultaneously serve multiple users over the same time-frequency resource, significantly increasing spectral efficiency through spatial multiplexing. Large antenna arrays also enable high-resolution angle estimation, making M-\ac{MIMO} fundamental for achieving high-accuracy 6G localization. At FR1 (and likely FR3), digital arrays with dedicated RF chains per antenna element provide accurate angle estimation and robust multipath separation in dense propagation environments, avoiding the time-consuming beam sweeping required by hybrid and analog arrays. However, challenges include high hardware complexity and power consumption from the large number of RF chains, intricate multipath propagation (e.g., multi-bounce reflections and diffraction) and limited bandwidth, resulting in poor range resolution \cite{thz_6G_2019}. At FR2 (and likely FR4), M-\ac{MIMO} accommodates denser antenna arrays under half-wavelength spacing and broader bandwidths, enabling narrow, high-gain beams that boost \ac{SNR} and improve delay and angular resolution. The beam squint effect at these frequencies may also offer additional spatial cues for localization by illuminating different spatial regions at varying frequencies.  However, localization with M-\ac{MIMO} at FR2 and FR4 suffers from hardware imperfections (e.g., power amplifier nonlinearity and phase noise), analog/hybrid array constraints limiting mobility support and angle estimation, and harsh propagation conditions including \ac{los} blockage and diffuse scattering \cite{6g_loc_sens_2024}. 

\subsubsection{Distributed MIMO}
\ac{dmimo} technology employs multiple geographically distributed \acp{ap}, coordinated to function as a unified system, providing improved \ac{SNR}, increased uniformity of service quality, and superior interference management compared to M-\ac{MIMO} systems. The benefits of \ac{dmimo} in 6G localization include extended coverage through spatial diversity and enhanced multipath resolvability via phase-coherent processing in rich propagation environments \cite{guo2024integrated}. At FR1 and FR3, \ac{dmimo} systems can operate in phase-coherent mode, which transforms the entire network into an extremely large sparse array, substantially improving angular resolution. Due to relatively small bandwidths available at FR1 and FR3, phase coherence among distributed \acp{ap} is crucial for high-accuracy localization as geometric information is conveyed through phase measurements \cite{guo2024integrated}. Conversely, small phase and positional misalignment can severely deteriorate localization performance, placing stringent requirements on synchronization and geometry calibration. Moving higher up in frequency, phase-coherence across distributed nodes becomes more challenging. Therefore, at FR2 and FR4, \ac{dmimo} typically shifts from phase-coherent to time-coherent mode and offers favorable delay resolution through large available bandwidths. Although time synchronization is less demanding than phase synchronization from both hardware and algorithmic perspectives, which benefits localization at FR2 and FR4, worsening hardware impairments can still degrade localization accuracy compared to that observed at FR1 and FR3.

\subsubsection{Reconfigurable Intelligent Surfaces}
\Acp{RIS} can dynamically control wireless signal propagation, enhancing coverage and signal quality in complex environments~\cite{chen2023riss}. \ac{RIS} technology includes various types, such as passive (low cost), active (high \ac{SNR}), and simultaneously transmitting and reflecting (STAR)-\ac{RIS} (wider coverage), each with unique operational characteristics. In all \ac{RIS} types, out-of-band control can be exploited to share geometric information, which can aid in coordinating multiple \ac{RIS} units and dynamically adjust their phase shifts. Acting as an extra anchor (i.e., a location reference node), a \ac{RIS} can significantly improve localization by offering additional reference points for positioning, especially in \ac{nlos} scenarios. However, calibration of the surveyed position and orientation of \ac{RIS}, non-ideal beam patterns, and variations in reflection coefficients are essential to achieve high-precision positioning.  In addition to RIS calibration, the key challenges to be addressed include the computational complexity of optimizing \ac{RIS} configurations in real-time and the high path loss caused by the \ac{RIS}'s multiplicative/cascaded double-fading channel.

\subsubsection{Sidelink Communications} By leveraging sidelink communication, nodes, such as vehicles and \acp{uav}, can share absolute or relative measurements to achieve higher positioning accuracy and robustness~\cite{liang2024toward}. This approach allows for both explicit cooperation (sharing exact location information) and implicit cooperation (using relative geometric measurements between nodes), enhancing localization accuracy. However, cooperative localization requires an increased demand for bandwidth and efficient resource allocation to handle the extra data exchange. Privacy and security are also concerns, as sharing location or relevant measurement data can expose sensitive information. Advanced solutions like federated learning can address these privacy issues, while decentralized scheduling can reduce the data load and energy consumption to ensure scalability.

\subsection{Non-terrestrial Networks}
\acp{NTN} come in many variants and can be broadly categorized into space-based  and aerial-based segments. In the former category, we count \ac{GEO} satellites, \ac{MEO} satellites, and \ac{LEO} satellites. In the latter category, we count \acp{HAPS} and \acp{uav}. In the following, we will discuss the technologies in each segment, highlighting their characteristics, opportunities, and challenges.

\subsubsection{Space-based \acp{NTN}}
Satellites share common traits such as high altitude above ground, susceptibility to atmospheric effects, and orbital state uncertainty. \ac{GEO} satellites orbit at $35,786$ km, \ac{MEO} satellites at $7,000$–$25,000$ km, and \ac{LEO} satellites at $300$–$1,500$ km above sea level. Their altitudes impact latency, path loss, beam footprint size, and \ac{los}/\ac{nlos} characteristics, influencing link budgets, \ac{SNR}, and susceptibility to jamming and spoofing. In addition, high satellite altitude results in poor positioning information gained from angle-based measurements. Hence, positioning with an independent satellite system requires access to multiple satellites to perform positioning, which might not be feasible in all scenarios.
Next, atmospheric effects include ionospheric (more prominent in S and L bands, similar to FR1) and tropospheric effects (more prominent in Ku/FR3 and Ka/FR2 bands). These effects, categorized as fast (e.g., scintillation) or slow (e.g., absorption), can distort signal phases and must be modeled, estimated, and compensated to avoid positioning errors. 
Finally, orbital state uncertainty arises from factors like gravitational forces, atmospheric drag, and solar radiation pressure, which disturb satellite trajectories and require correction to enhance positioning accuracy. 

A key distinction among these anchors is orbital mobility. \ac{GEO} satellites, geosynchronous with Earth, experience minimal Doppler effects. On the other hand, \ac{MEO} satellites, with approximately 12-hour orbital periods, exhibit moderate Doppler shifts, while \ac{LEO} satellites, which typically take between 90 minutes and 2 hours to complete one full orbit, experience significant shifts. These Doppler shifts aid positioning but necessitate advanced estimation techniques.
Differences in satellite footprints also matter. \ac{LEO} systems require mega-constellations for global coverage, \ac{MEO} constellations need fewer satellites, and \ac{GEO} satellites, fixed above the equator, provide wider coverage. \ac{LEO} and \ac{MEO} constellations, offering diverse geographic observations, improve \ac{GDoP} and positioning accuracy but introduce coordination challenges that will be discussed later. 
Lastly, satellites are equipped with clocks that vary in quality and stability. For instance, \ac{GEO} and \ac{MEO} satellites, being less numerous and typically heavier, are equipped with atomic clocks, which are far more stable then the clocks in the more abundant and smaller \ac{LEO} satellites. Hence, extra attention must be directed towards modeling of these clocks and estimating their biases to avoid loss in localization accuracy \cite{9193893}.

\subsubsection{Aerial-based \acp{NTN}} 
Aerial nodes, encompassing \acp{HAPS} and \acp{uav}, operate at lower altitudes than space-based systems, typically between $20-50$ km for \acp{HAPS} and a few hundred meters to several kilometers for \acp{uav}, depending on the local regulations. Their proximity to Earth results in lower signal delays, transmission power, and path losses compared to satellites. \acp{HAPS} offer quasi-stationary wide-area coverage, similar to \ac{GEO} satellites, making them suitable for remote and rural areas, while \acp{uav} provide flexible and dynamic coverage in urban or emergency scenarios. Both systems primarily contend with tropospheric effects like rain attenuation and fog, as well as state uncertainties due to wind drift, affecting positioning accuracy. Furthermore, mobility varies between these systems. For instance, \acp{HAPS} are relatively stable, experiencing minimal Doppler shifts, whereas \acp{uav} can introduce moderate Doppler effects due to their rapid movements, providing additional positioning information and adding complexity to using them as anchors. Despite these challenges, \acp{HAPS} and \acp{uav} enhance localization performance by increasing the number of anchor points, enhancing the vertical \ac{GDoP}, offering adaptable coverage, and bridging coverage gaps. However, they also introduce complexities in terms of real-time coordination, power management, interference handling, and path planning (in the case of \acp{uav}), which must be carefully addressed to realize their full localization potential \cite{9193893}.

\subsection{Frequency Dependencies}
Table \ref{tab:localization_pros_cons} summarizes the interaction of both terrestrial and non-terrestrial segments and the various envisioned 6G frequency ranges. The table shows that, in general, \ac{TN} and \ac{NTN} networks operating at lower frequency ranges are more mature, have higher coverage, are less affected by hardware impairments, environmental effects, and user mobility, and can achieve low to medium levels of positioning accuracy. On the other hand, operating at higher frequencies holds the potential of achieving higher positioning accuracy but at the cost of complexity of addressing high signal attenuation, \ac{los} blockage, environmental effects, \ac{ue} mobility issues, and hardware impairments. Hence, more research work needs be done to tackle these aspects in order to achieve the highest potential of these 6G enabler technologies.

%\cellcolor{green!25}\textbf{Pros} & \cellcolor{red!25}\textbf{Cons} & \cellcolor{green!25}\textbf{Pros} & \cellcolor{red!25}\textbf{Cons} 
%{black!15} {blue!15}
% FRs FFE699 F7914D A9D18E BB99DD 
% Pros and cons D8F5A2 C85C53
% TN and NTN FFF5D8 BDD7EE
\begin{table*}[ht!]
    \caption{Localization Pros and Cons Across Different Frequency Ranges}
    \centering
    \renewcommand{\arraystretch}{1.2}
    \begin{tabular}{|>{\centering\arraybackslash}m{3.3cm}|m{3.25cm}|m{3.25cm}|m{3.25cm}|m{3.25cm}|}
        \hline
            \multirow{2}{*}{\textbf{\shortstack{Frequency \\Range}}} & \multicolumn{2}{c|}{\cellcolor[HTML]{FFF5D8}\textbf{TN}} & \multicolumn{2}{c|}{\cellcolor[HTML]{BDD7EE}\textbf{NTN}} \\ \cline{2-5}
        & \cellcolor[HTML]{D8F5A2}\textbf{Pros} & 
          \cellcolor[HTML]{C85C53}\textbf{Cons} & 
          \cellcolor[HTML]{D8F5A2}\textbf{Pros} & 
          \cellcolor[HTML]{C85C53}\textbf{Cons} 
        \\ \hline
        \cellcolor[HTML]{FFFFFF}\textbf{\shortstack{FR1: \textless 7.125 GHz\\ L and S bands}} 
        & Mature technologies, phase-coherence exploitation in D-MIMO
        & Poor localization performance due to limited bandwidth and need for several visible BSs
        & Wide coverage, penetration capabilities 
        & Lower range accuracy, ionospheric effects
        \\ \hline

        \cellcolor[HTML]{FFFFFF}\textbf{\shortstack{FR3: 7.125--24.25  GHz\\ X, Ku, and K bands}}
        & Improved localization flexibility due to wider spectrum and availability of angle-based measurements
        & Fragmented spectrum and need for several visible BSs %Unexplored Less developed ecosystem
        & Balanced localization performance, low interference risks
        & Intermediate accuracy, tropospheric effects
        \\ \hline

        \cellcolor[HTML]{FFFFFF}\textbf{\shortstack{FR2: 24.25--52.6/57--71 GHz\\ Ka and V bands }} 
        & High capacity for accurate localization, low latency positioning%. , beam squint, dense deployment
        & Limited coverage in obstructed areas, higher cost of dense deployments, impairments, mobility limitations
        & High-precision localization and intermediate coverage
        & High signal attenuation, dependence on \ac{los} links, tropospheric effects
        \\ \hline

        \cellcolor[HTML]{FFFFFF}\textbf{\shortstack{FR4: 90--300  GHz\\ sub-THz bands}} 
        & Massive potential for high-accuracy positioning, minimal latency%, beam squint, dense deployment
        & LoS limitations, high cost of infrastructure,  impairments, less mature technology, mobility limitations
        & Ultra-high precision localization, high security and privacy
        & Extreme path loss, sensitive to weather and environment, misalignment, less mature technology
        \\ \hline
    \end{tabular}
    \label{tab:localization_pros_cons}
\end{table*}

%%%%%%%%%%%%%%%%%%%%%%%%%%%%%%%%%%%%%%%%%%%%%%%%%%%%%%%%%%%%%%%
\section{Standardization Activities and Industrial Advancements}
While 6G standardization is expected to commence with 3GPP's Release 20 by 2025, the groundwork for \ac{TN} and \ac{NTN} localization enablers has already been laid in 5G and 5G-Advanced standards (Releases 16 to 19). Additionally, proprietary industrial solutions, such as Starlink, have emerged over the past decade, offering valuable insights and lessons for the development of the next-generation cellular \ac{NTN}. Hence, this section explores these standardization efforts and industrial advancements in \ac{TN} and \ac{NTN} positioning.

\subsection{3GPP Standardization Activities}
Although 3GPP positioning standardization has focused on TN over the past decades, NTN standardization started to gain traction and is envisioned to continue in 6G. In this section, we review 3GPP's 5G positioning standards in TN and NTN, and our vision for 3GPP's 6G TN-NTN integration standardization.
%3GPP standardization efforts have initially focused on TN positioning and may be extended to NTN positioning in the future. Hence, further activities on TN-NTN integration and exploration of new frequency spectrum are expected in the near future.

\subsubsection{5G TN Standardization}  Positioning services have been supported in all generations of cellular networks, starting from supporting emergency services and evolving the 5G system to also support commercial services with tighter positioning requirements. In terms of technology, one of the major innovations occurred in the 4G era, where dedicated reference signals for positioning, such as \ac{DL} \ac{PRS} and \ac{UL} \ac{SRS}, were introduced. These came along with various positioning techniques, such as the \ac{DL}-\ac{TDOA}, \ac{UL}-\ac{TDOA}, and \ac{e-CID} \cite{1G_5G_survey_2018}. 
In 5G, further enhancements were introduced, such as the usage of higher frequency (FR2) and M-MIMO to enable beamforming-based transmission and reception, which together with other enhancements, allowed  new target requirements of 20~cm in accuracy and less than 10~ms in latency to be achieved. Furthermore, challenging scenarios where \ac{nlos} becomes dominant were also investigated. These resulted in enhancements and new positioning techniques being introduced in 3GPP Release 17 \cite{wang_recent_2023}. 
In 3GPP Release~18, even further positioning enhancements, such as the use of carrier phase measurements, and device-to-device positioning were introduced \cite{wang_recent_2023}. 
In the current 3GPP Release~19, a new feature on the usage of artificial intelligence/machine learning (AI/ML) for positioning enhancements is being specified. 

\subsubsection{5G NTN Standardization} \ac{NTN} was first studied in 3GPP Release 16 and then specified in Release~17. Although positioning procedures and solutions are still primarily developed for the \ac{TN}, \ac{NTN} still continuously evolves with new features for NR and \ac{IoT} devices. Among these features, there is a limited positioning operation in \ac{NTN} that was introduced in 3GPP Rel-18 with the purpose of supporting network-verified \ac{ue} location with extremely coarse positioning accuracy (i.e., of the order of 10 km)~\cite{wang_recent_2023}. This is to ensure that the \ac{ue} is connected to the appropriate core network, particularly for a \ac{ue} close to country borders. For this purpose, the method uses multiple satellite-UE \ac{RTT} measurements at different time instances as described in \cite{10355106}. 

\subsubsection{6G NTN and TN Standardization} In the 6G time frame, \ac{NTN} positioning is expected to have tighter requirements to support more use cases. The legacy positioning measurements in \ac{TN} positioning, such as angle-based measurement (e.g., \ac{aoa} and \ac{aod}) and timing-based measurement (e.g., \ac{DL}-\ac{TDOA}, UL-\ac{TDOA}, and multi-\ac{RTT}), can be extended to \ac{NTN} positioning. In \ac{NTN} positioning, deployment scenarios where the \ac{ue} sees either one or multiple satellites need to be considered. This significantly affects how the positioning measurements are performed and on how the final location is estimated. The positioning measurement/estimation can differ depending on the aforementioned deployment scenarios.  NTN positioning with only one satellite can be challenging in order to achieve accurate positioning. At the start of 6G, there is an opportunity to investigate new reference signals/waveforms for positioning. Such new signals could be adopted, especially when they prove beneficial in comparison to the legacy \ac{PRS} or \ac{SRS}. The network architecture to support \ac{NTN} positioning is expected to be developed based on the legacy \ac{TN} positioning architecture, by involving the \ac{LMF}. However, it is expected that there will be some enhancements through adding new features/functions and also in the signaling mechanism between network nodes. Although \ac{TN} and \ac{NTN} positioning have operated independently and have been used for different purposes in 5G, we envision that 6G \ac{TN} and \ac{NTN} will be co-designed from the start. This will facilitate a common signal design and network architecture, enabling a smooth integration of \ac{TN} and \ac{NTN}.

\subsection{Industrial Advancements}
Industrial players, such as SpaceX, play a crucial role in exploring alternative and complementary solutions that have not been addressed by 3GPP standardization. In particular, the use of \ac{LEO} satellites for \ac{PNT} has attracted significant attention in recent years. These efforts include the deployment of dedicated \ac{LEO} constellations for \ac{PNT} services, as well as leveraging \ac{SoOP} from constellations originally designed for communication purposes. In this section, we discuss advancements in both dedicated and opportunistic \acp{NTN}.

\subsubsection{LEO PNT with Dedicated Systems} 
\ac{LEO} constellations specifically designed for \ac{PNT} are being developed as complementary systems to GNSS or as standalone alternatives. Initiatives such as the European Space Agency's efforts to develop a dedicated \ac{LEO}-\ac{PNT} constellation highlight this trend. Companies like TrustPoint, Xona Space Systems, Geely, and Future Navigation are actively developing their own \ac{LEO} satellite constellations, consisting of 288, 258, 240, and 160 satellites respectively, to deliver high-accuracy \ac{PNT} services. These systems are required to transmit ephemeris data, clock bias, and drift corrections, and may include atmospheric effects, providing the essential information for precise \ac{PNT} solutions. Different signal structures are under investigation for \ac{LEO}-\ac{PNT} applications, including \ac{OFDM} signals, direct-sequence spread spectrum (DSSS) signals, and chirp spread spectrum (CSS) signals, each with specific advantages and challenges~\cite{10714965}. Direct-sequence spread spectrum (DSSS) signals, commonly employed in GNSS systems, can face challenges in acquisition and tracking due to the rapid motion of LEO satellites, necessitating modifications from standard GNSS receivers. In contrast, CSS signals can avoid the two-dimensional Doppler-delay search required for acquisition in the GNSS architecture, enabling lower complexity solutions in scenarios with large Doppler shifts. However, further investigation is required to achieve accurate ranging, access multiple satellites, and enable data transmission using CSS signals for \ac{LEO}-\ac{PNT}. Currently, many aspects of these dedicated \ac{LEO}-\ac{PNT} systems are still under development, with ongoing efforts to refine technology and deploy infrastructure.

\subsubsection{LEO PNT via SoOP}
\ac{LEO} satellites, originally designed for non-navigation purposes, can also serve as valuable resources for localization by opportunistically utilizing their signals. Existing \ac{LEO} constellations suitable for \ac{PNT} via \ac{SoOP} include Starlink, Orbcomm, Argos, Iridium, Globalstar, and others, each operating with distinct frequency bands and modulation schemes~\cite{10542356}. Currently, thousands of satellites from multiple operators are in orbit, with tens of thousands anticipated in the near future. A key advantage of \ac{SoOP} is its ability to utilize a wide variety of ambient satellite signals, increasing signal diversity and maximizing resource efficiency. However, the absence of signal specifications, often due to business security or privacy concerns, introduces considerable challenges in signal processing and synchronization. Currently, efforts are underway to develop advanced signal processing techniques and receiver architectures capable of extracting key observations for positioning applications, including Doppler shift, carrier phase, and pseudo-range. This can be achieved by leveraging the inherent characteristics of these signals, along with techniques such as blind beacon estimation, machine learning-based signal processing, and other advanced methods~\cite{10542356}.

\begin{figure*}
    \centering
    \includegraphics[trim=0pt 0pt 0pt 27pt,clip,width=1\linewidth]{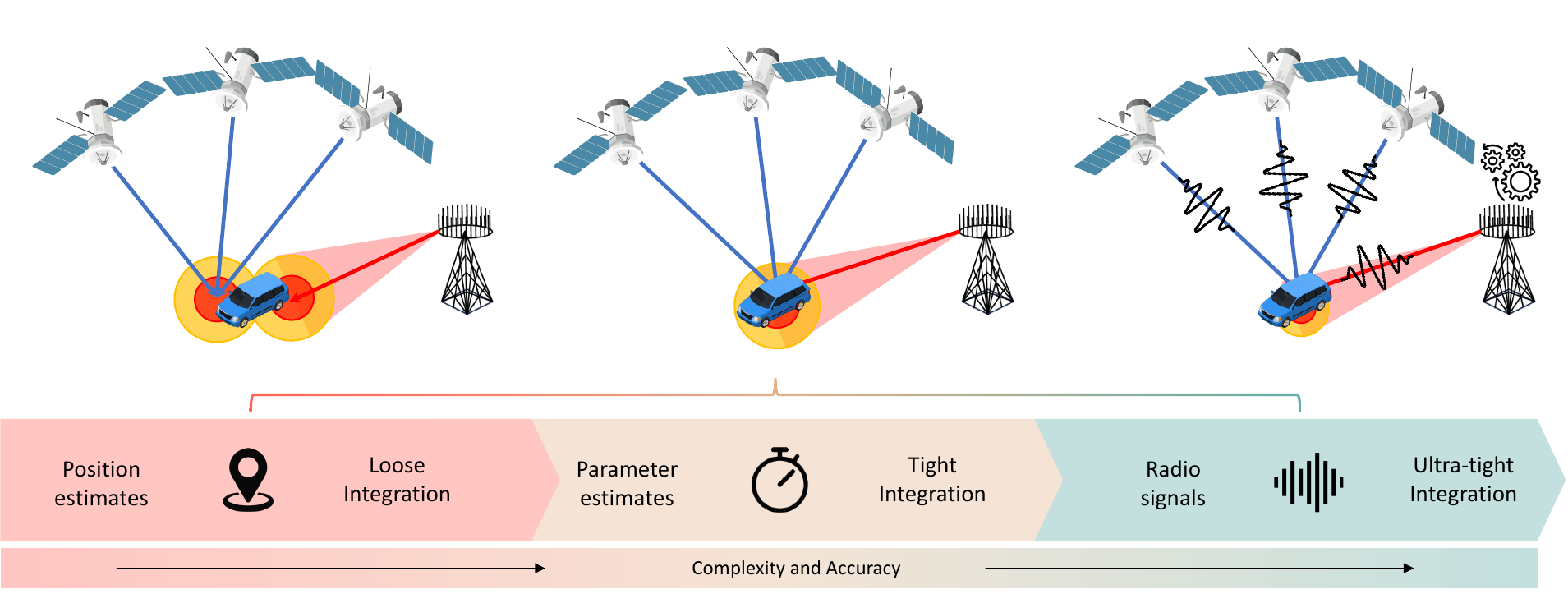}
    \caption{Levels of TN-NTN integration. Loose integration (left) combines the final position estimates from each network. Tight integration (middle) fuses parameter estimates such as range and angle measurements to determine the UE's position. Ultra-tight integration (right) co-designs and processes raw I/Q samples from both networks for ultimate enhancement of positioning performance. Higher integration levels offer improved accuracy but come with increased complexity.}
    \label{fig:enter-label}
\end{figure*}

%%%%%%%%%%%%%%%%%%%%%%%%%%%%%%%%%%%%%%%%%%%%%%%%%%%%%%%%%%%%%%%
\section{Integration Challenges, Opportunities, and Case Studies}
\begin{figure}
    \centering
    \includegraphics[width=1\linewidth]{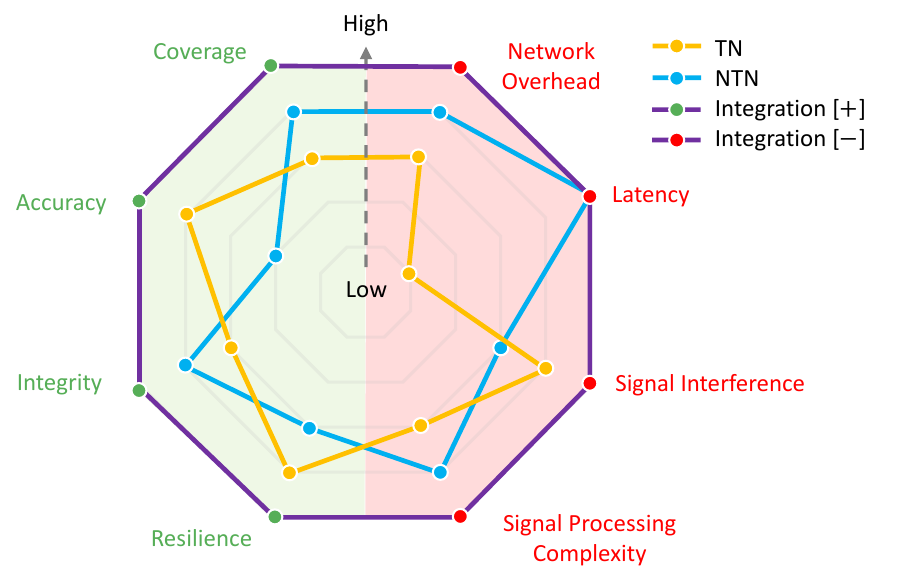}
    \caption{Evaluation of characteristics of TN, NTN, and their integration for localization purposes. The characteristics on the left-hand side are positive  (further from the center is better) and the characteristics on the right-hand side are negative (closer to the center is better).}
    \label{6G_integration_ntn_tn}
\end{figure}

While both \ac{TN} and \ac{NTN} offer advanced localization capabilities, their integration remains largely unexplored. Combining these networks promises higher accuracy and a seamless, globally unified localization service. Integration can occur at various stages of the localization pipeline, which includes designing the communication system, estimating geometric channel parameters, and fusing those parameters to determine the user's position. Hence, integration is typically categorized into three levels: loose, tight, and ultra-tight, as shown in Fig.~\ref{fig:enter-label}. 
\textit{Loose integration} combines the final position estimates from each system. For instance, it requires independent multilateration from at least four non-terrestrial anchors, along with a positioning solution from one or more terrestrial BSs, which are then fused to enhance accuracy. \textit{Tight integration}, by contrast, operates on geometric measurements, removing the need for independent multilateration from multiple non-terrestrial anchors. It can fuse a single range measurement from an NTN anchor with range and angle measurements from a BS or an RIS, offering greater flexibility and accuracy at the cost of increased system complexity. \textit{Ultra-tight integration} takes this further by modifying both networks at earlier stages. These modifications, ranging from joint resource allocation to unified physical-layer design, significantly boost performance but also add complexity. 
Although every integration schemes has its pros and cons, all of them share a set of challenges to be solved and opportunities to be reaped. Hence, this section outlines the key challenges and opportunities of \ac{TN}-\ac{NTN} integration, summarized in Fig.~\ref{6G_integration_ntn_tn}, and explores its potential through two tight integration case studies to highlight the opportunities in this domain.

% This integration can occur at different levels, including the steps of position estimation, parameter estimation, and radio signal design, as shown in Fig.~\ref{fig:enter-label}. The data fusion at positioning and parameter estimation stages is expected to impose higher complexity on algorithms without altering hardware. In contrast, integration at the radio signal level typically requires a unified physical-layer design, enhancing positioning performance at the cost of increased hardware complexity. Nevertheless, integration at any level presents shared challenges and opportunities. This section first outlines several key challenges and opportunities of \ac{TN}-\ac{NTN} integration, summarized in Fig.~\ref{6G_integration_ntn_tn}. Then, it examines the potential of such hybrid systems through two case studies, further highlighting the opportunities within this domain.

\subsection{Challenges}
Integrating TN and NTN for localization presents several technical challenges, including increased network overhead, higher processing latency, stronger signal interference, and greater signal processing complexity, as highlighted in red in Fig.~\ref{6G_integration_ntn_tn}. These challenges need to be addressed before we are able to reap the benefits of the TN-NTN integration.

\subsubsection{Network Overhead} The integration of TN and NTN requires frequent information exchange among terrestrial base stations, gateways, non-terrestrial anchors, and user terminals, leading to increased signaling overhead. This is particularly evident in managing handovers, synchronization, and control signaling across diverse links. Such overheads will be exacerbated as the TN-NTN integration becomes tighter.

\subsubsection{Latency} Compared to \ac{TN}, the long propagation distances inherent to \ac{NTN} links introduce significant delays, which can impair time-sensitive applications such as vehicular networks and autonomous systems. Such latency is expected to increase due to the extra signaling needed for integration.

\subsubsection{Signal Interference} The coexistence of terrestrial and non-terrestrial links creates complex interference scenarios, which can arise from overlapping frequency bands, multipath propagation, or inter-satellite, \ac{HAPS}, and \ac{uav} links. Managing such interference in a dynamic environment adds further complexity to system design and operation. 

\subsubsection{Signal Processing} Achieving high localization accuracy in integrated systems requires advanced signal processing algorithms to address challenges, such as varying observations, non-stationary fading, fluctuating SNR levels, and coupled external factors (e.g., atmospheric effects, hardware impairments, mobility, and anchor state uncertainty). These factors collectively contribute to a significant increase in computational complexity.

\subsection{Opportunities}
Integrated \ac{TN}-\ac{NTN} systems will offer significant opportunities in terms of coverage, accuracy, integrity, and resilience, as highlighted in green in Fig.~\ref{6G_integration_ntn_tn}. 

\subsubsection{Coverage} Both \ac{TN} and NTN have coverage limitations. \ac{NTN} struggles in urban canyons, dense indoor areas, and regions with signal blockage, while \ac{TN} faces challenges in remote or rural areas. The integration of \ac{TN} and \ac{NTN} overcomes these limitations by combining \ac{NTN}'s global coverage with \ac{TN}'s regional reach, ensuring seamless localization across all environments and providing ubiquitous coverage. 

\subsubsection{Accuracy} \ac{TN}-\ac{NTN} integration can enhance localization accuracy for the following two reasons: (i) it leverages multi-source data fusion, and (ii) the dispersed localization anchors in ground, air, and space environments significantly improve the \ac{GDoP}, leading to more accurate localization performance.

\subsubsection{Integrity} \ac{TN}-\ac{NTN} systems offer higher localization integrity, i.e., enhanced trustworthiness and reliability, by cross-validating information from multiple sources. This is critical for safety-sensitive applications such as aviation, maritime navigation, and autonomous driving. 

\subsubsection{Resilience} \ac{TN}-\ac{NTN} integration improves resilience against jamming and spoofing by leveraging signal diversity and redundancy across terrestrial and non-terrestrial networks. Non-terrestrial anchors, being less vulnerable to ground-based jamming, provide an additional layer of robustness, while advanced signal processing techniques, like jammer localization, enable interference suppression and anomaly detection.
%The increased geometrical diversity of the integrated localization systems enhances robustness against jamming and spoofing attacks in comparison to independent TN and NTN systems. This can be achieved by locating the 

\begin{figure}[t]
    \centering
% \definecolor{mycolor1}{rgb}{0.00000,1.00000,1.00000}%
\definecolor{mycolor1}{rgb}{0.00000,0.44700,0.74100}%
\definecolor{mycolor2}{rgb}{0.85000,0.32500,0.09800}%
\definecolor{mycolor3}{rgb}{0.92900,0.69400,0.12500}%
\definecolor{mycolor4}{rgb}{0.49412,0.18431,0.55686}%
\definecolor{mycolor5}{rgb}{0.49400,0.18400,0.55600}%
\begin{tikzpicture}
\begin{axis}[%
width=0.8\columnwidth,
height=2in,
at={(0pt,0pt)},
scale only axis,
xmin=0,
xmax=50,
xlabel style={font=\color{white!15!black}, align=center},
xlabel={BS transmit power [dBm]},
ymode=log,
ymin=5e-5,
ymax=1e2,
yminorticks=true,
ylabel style={font=\color{white!15!black}},
ylabel={CRLB [m / deg]},
axis background/.style={fill=white},
xmajorgrids,
ymajorgrids,
yminorgrids,
legend style={legend cell align=left, align=left, draw=white!15!black, legend columns=2}
% legend style={
%         at={(0.5,-0.2)}, % Positioning the legend
%         anchor=north, % Legend's anchor point
%         legend columns=2, % Two columns
%     },
%     legend cell align={left}, % Align legend entries
%     width=10cm, % Adjust width as needed
%     height=6cm % Adjust height as needed
]

\addplot [color=mycolor3, line width=1.5pt, mark=triangle*, mark options={solid, mycolor3}]
  table[row sep=crcr]{%
0	0.29023267705864\\
2	0.29023267705864\\
4	0.29023267705864\\
6	0.29023267705864\\
8	0.29023267705864\\
10	0.29023267705864\\
12	0.29023267705864\\
14	0.29023267705864\\
16	0.29023267705864\\
18	0.29023267705864\\
20	0.29023267705864\\
22	0.29023267705864\\
24	0.29023267705864\\
26	0.29023267705864\\
28	0.29023267705864\\
30	0.29023267705864\\
32	0.29023267705864\\
34	0.29023267705864\\
36	0.29023267705864\\
38	0.29023267705864\\
40	0.29023267705864\\
42	0.29023267705864\\
44	0.29023267705864\\
46	0.29023267705864\\
48	0.29023267705864\\
50	0.29023267705864\\
};
\addlegendentry{$\tau_{\text{Sat},{45\text{dBm}}}$}% = 0 [dBm]}

\addplot [color=black, line width=1.5pt, mark=triangle, mark options={solid, black}]
  table[row sep=crcr]{%
0	1.81683011821701\\
2	1.45372887794296\\
4	1.16792038438605\\
6	0.944070035226042\\
8	0.770045880686829\\
10	0.636205637987467\\
12	0.53479268040842\\
14	0.4594281754392\\
16	0.40471875308825\\
18	0.366015726160203\\
20	0.339332001650064\\
22	0.32135769333894\\
24	0.309480000633669\\
26	0.301745211828576\\
28	0.296761178655915\\
30	0.29357293251601\\
32	0.291543346108729\\
34	0.290255461700119\\
36	0.289439912944871\\
38	0.288924152019491\\
40	0.288598254891574\\
42	0.288392438154768\\
44	0.28826250104455\\
46	0.288180484540207\\
48	0.288128725349572\\
50	0.2880960587149\\
};
\addlegendentry{$\boldsymbol{p}_{45\text{dBm}}$}%}$ = 0 [dBm]}

\addplot [color=mycolor3, dashed, line width=1.5pt, mark=square*, mark options={solid, mycolor3}]
  table[row sep=crcr]{%
0	0.029023267705864\\
2	0.029023267705864\\
4	0.029023267705864\\
6	0.029023267705864\\
8	0.029023267705864\\
10	0.029023267705864\\
12	0.029023267705864\\
14	0.029023267705864\\
16	0.029023267705864\\
18	0.029023267705864\\
20	0.029023267705864\\
22	0.029023267705864\\
24	0.029023267705864\\
26	0.029023267705864\\
28	0.029023267705864\\
30	0.029023267705864\\
32	0.029023267705864\\
34	0.029023267705864\\
36	0.029023267705864\\
38	0.029023267705864\\
40	0.029023267705864\\
42	0.029023267705864\\
44	0.029023267705864\\
46	0.029023267705864\\
48	0.029023267705864\\
50	0.029023267705864\\
};
\addlegendentry{$\tau_{\text{Sat},{65\text{dBm}}}$}% = 40 [dBm]}

\addplot [color=black, dashed, line width=1.5pt, mark=square, mark options={solid, black}]
  table[row sep=crcr]{%
0	1.79408310507112\\
2	1.42519828796184\\
4	1.1322104635225\\
6	0.899516948620431\\
8	0.714725938948246\\
10	0.567996587357345\\
12	0.451514917206412\\
14	0.35907764874805\\
16	0.285761751368438\\
18	0.227662076860046\\
20	0.18168301182168\\
22	0.145372887794289\\
24	0.116792038438663\\
26	0.0944070035226516\\
28	0.077004588068711\\
30	0.0636205637987065\\
32	0.0534792680406733\\
34	0.0459428175440275\\
36	0.0404718753090257\\
38	0.0366015726162168\\
40	0.033933200165354\\
42	0.0321357693337822\\
44	0.030948000064678\\
46	0.0301745211845055\\
48	0.0296761178634879\\
50	0.0293572932424177\\
};
\addlegendentry{$\boldsymbol{p}_{65\text{dBm}}$}%}$ = 40 [dBm]}

\addplot [color=mycolor2, line width=1.5pt, mark=+, mark options={solid, mycolor2}]
  table[row sep=crcr]{%
0	0.39146027830195\\
2	0.310947951828264\\
4	0.246994737666876\\
6	0.196194893957116\\
8	0.155843143778874\\
10	0.123790609291755\\
12	0.0983303761541631\\
14	0.078106594110311\\
16	0.0620422730199693\\
18	0.0492819292062341\\
20	0.0391460278301949\\
22	0.0310947951828264\\
24	0.0246994737666875\\
26	0.0196194893957116\\
28	0.0155843143778874\\
30	0.0123790609291755\\
32	0.00983303761541631\\
34	0.0078106594110311\\
36	0.00620422730199693\\
38	0.00492819292062342\\
40	0.0039146027830195\\
42	0.00310947951828263\\
44	0.00246994737666876\\
46	0.00196194893957116\\
48	0.00155843143778874\\
50	0.00123790609291755\\
};
\addlegendentry{$\theta_\text{BS}^\text{az}$}

\addplot [color=mycolor1, line width=1.5pt, mark=+, mark options={solid, mycolor1}]
  table[row sep=crcr]{%
0	0.258110926204848\\
2	0.205024796375346\\
4	0.162856984579534\\
6	0.129361901073581\\
8	0.102755810520355\\
10	0.081621841578296\\
12	0.0648345333358328\\
14	0.0514999004138247\\
16	0.0409078249841896\\
18	0.0324942404061012\\
20	0.0258110926204848\\
22	0.0205024796375346\\
24	0.0162856984579534\\
26	0.0129361901073581\\
28	0.0102755810520355\\
30	0.0081621841578296\\
32	0.00648345333358328\\
34	0.00514999004138246\\
36	0.00409078249841896\\
38	0.00324942404061012\\
40	0.00258110926204848\\
42	0.00205024796375346\\
44	0.00162856984579534\\
46	0.00129361901073581\\
48	0.00102755810520355\\
50	0.000816218415782961\\
};
\addlegendentry{$\theta_\text{BS}^\text{el}$}

\addplot [color=mycolor4, line width=1.5pt, mark=o, mark options={solid, mycolor4}]
  table[row sep=crcr]{%
0	0.0158396409221701\\
2	0.0125818740123739\\
4	0.00999413777377222\\
6	0.00793862581543175\\
8	0.00630587463010851\\
10	0.00500893426332674\\
12	0.00397873791123829\\
14	0.00316042386146448\\
16	0.00251041390685759\\
18	0.00199409264706148\\
20	0.001583964092217\\
22	0.00125818740123739\\
24	0.000999413777377223\\
26	0.00079386258154318\\
28	0.000630587463010851\\
30	0.000500893426332674\\
32	0.000397873791123829\\
34	0.000316042386146448\\
36	0.000251041390685759\\
38	0.000199409264706147\\
40	0.000158396409221701\\
42	0.000125818740123739\\
44	9.99413777377224e-05\\
46	7.93862581543175e-05\\
48	6.30587463010851e-05\\
50	5.00893426332672e-05\\
};
\addlegendentry{$\tau_{\text{BS}}$}

\end{axis}
\end{tikzpicture}%
    \vspace{-3em}
        \caption{Illustration of localization performance in a single-LEO single-BS scenario. The \ac{CRLB} of delays and position estimates are in meters and the \ac{CRLB} of angles are in degrees. Two satellite transmission powers were tested, 45 dBm and 65 dBm, illustrated in solid lines with triangle marks and dotted lines with square marks, respectively. The setup operates at 2 GHz carrier frequency, 50 MHz bandwidth, single antenna at the UE (80 km/h) and the satellite, 2x2 array at the BS, 15 dB noise figure, 4 OFDM symbols, 1200 km LEO satellite altitude, 250 m UE-BS distance, and UE-network clock bias and carrier frequency offset (CFO) assumptions.}
    \label{fig:CS1}
    \vspace{-1em}
\end{figure}
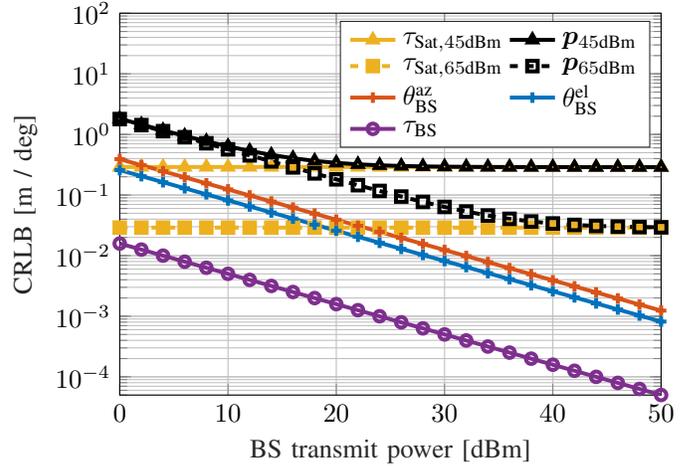

\subsection{Integration Case Studies}
In this section, we use two case studies to illustrate the potential of \ac{TN}-\ac{NTN} integration in localization. Both case studies are examples of tight TN-NTN integration (i.e., integration on the measurement level). However, there are slight differences when it comes to their positioning accuracy due to the different setups and channel models used.

\subsubsection{Single-BS-Single-LEO Localization}
The first case study explores the integration of a terrestrial \ac{bs} with a single \ac{LEO} satellite to localize a vehicular user in an urban scenario. This case study focuses on the effectiveness of arbitrarily increasing the \ac{bs}'s transmission power on the localization performance without coordination with its NTN counterpart. %in scenarios where \ac{CFO}, clock bias, and severe Doppler effects are present. In this scenario, %the \ac{LEO} satellite and the \ac{bs} simultaneously transmit 64 \ac{OFDM} symbols on orthogonal frequency resources that have an aggregate bandwidth of 100 MHz, operating in the 2 GHz band of FR1. 
In this scenario, two \ac{LEO} transmission power levels were tested while the \ac{bs} transmission power was varied from 0 dBm to 50 dBm. The simulation results, shown in Fig.~\ref{fig:CS1}, present the \ac{CRLB} on the \ac{bs}'s delay and angle estimation, the \ac{LEO}'s delay estimation, and the user's position estimation. The results show that increasing the \ac{bs}'s transmission power enhances \ac{bs}-based measurements, leading to improved positioning accuracy. However, this improvement continues only up to a certain point, beyond which the positioning accuracy saturates, constrained by the quality of satellite-based delay estimates. This is further verified as the saturation point was lowered when using higher transmission power at the satellite's side. This study highlights the need for proper coordination between \ac{NTN} and \ac{TN} to avoid wasting valuable resources without achieving significant improvements in positioning performance.

\begin{figure*}[t]
  \centering
  \includegraphics[width=0.99\linewidth]{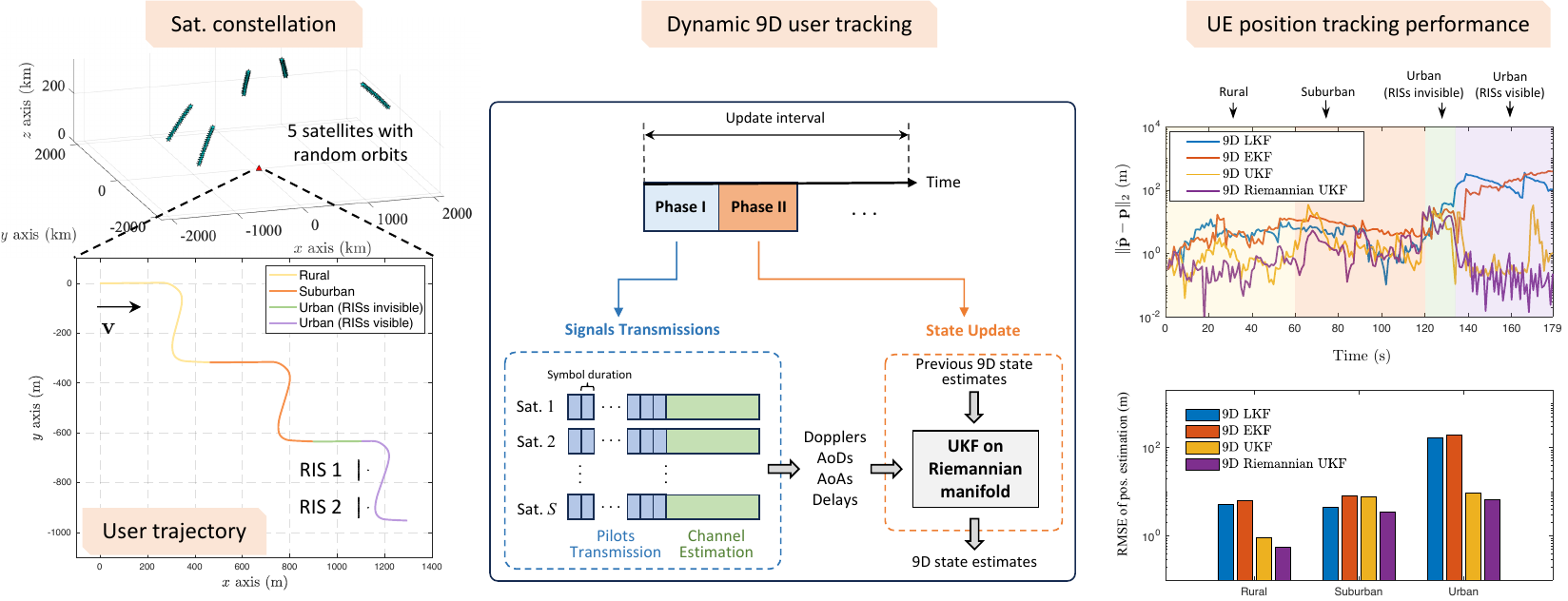}
  \vspace{-1em}
  \caption{ 
      Illustration and performance evaluation of a LEO- and \ac{RIS}-empowered 9D user tracking system. Based on the estimated channel parameters---such as Doppler shifts, \acp{aod}, \acp{aoa}, and channel delays---derived from the received pilot signals, we filter and update the \ac{3d} position, \ac{3d} velocity, and \ac{3d} orientation of the \ac{ue} using a designed unscented Kalman filter (UKF) on a Riemannian manifold~\cite{Zheng2024LEO}, benchmarked against the classical linearized Kalman filter (LKF), extended Kalman filter (EKF), and UKF. The performance has been comprehensively evaluated across various environments.}
  \label{fig:CS_LEORIS}
  \vspace{-0.7em}
\end{figure*}

\subsubsection{Multi-LEO- and Multi-RIS-empowered User Tracking}
In the second case study, we investigate a hybrid system that integrates \ac{LEO} satellites and terrestrial \acp{RIS} for user-tracking applications. Fig.~\ref{fig:CS_LEORIS} illustrates a cooperative \ac{DL} framework to coordinate different satellite transmissions and \ac{RIS} reflections. By exploiting the acquired channel parameters and the \ac{ue}'s motion dynamics, a tracking algorithm, proposed in~\cite{Zheng2024LEO}, enables comprehensive tracking of the \ac{3d} position, velocity, and orientation of the \ac{ue}. The algorithm is based on a UKF and Riemannian manifold theory to address inherent challenges such as nonlinear observation models, constrained unknown states, and time-varying observation uncertainties. From Fig.~\ref{fig:CS_LEORIS}, we observe that the considered system gains a significant performance improvement in the \ac{RIS}-visible areas, compared to \ac{RIS}-invisible areas. This suggests that integrating \ac{RIS} with \ac{LEO} satellites holds substantial potential for enhancing user tracking performance, with appropriate signal-processing algorithms.

%%%%%%%%%%%%%%%%%%%%%%%%%%%%%%%%%%%%%%%%%%%%%%%%%%%%%%%%%%%%%%%
\section{Conclusions and Outlook}
% The integration of terrestrial and non-terrestrial networks offers a promising yet challenging path toward high-precision 6G localization. In this article, we highlighted the strengths and weaknesses of 6G enablers in both segments. We have shown that by leveraging the complementary advantages of \ac{NTN} and \ac{TN}, 6G systems will provide seamless localization even in complex environments, supporting a broad range of applications. However, significant challenges remain, including network interoperability across heterogeneous platforms, high network overhead, managing Doppler effects, ensuring synchronization between \ac{TN} and \ac{NTN}, and mitigating overlapping sources of errors and hardware imperfections. Addressing these issues will require innovative signal processing and networking solutions. Furthermore, as standardization and industrial solutions continue to evolve, they must do so symbiotically to ensure efficient collaboration between the diverse \ac{TN} and \ac{NTN} enablers. Finally, the following questions still need an answer. What will be the best combination of integrated enablers for a given scenario? And what would be the suitable frequency of operation and resource allocation in that case?

The integration of terrestrial and non-terrestrial networks offers a promising yet challenging path toward high-precision 6G localization. In this article, we examined the strengths and weaknesses of 6G enablers in both segments, demonstrating that by leveraging the complementary advantages of \ac{TN} and \ac{NTN}, 6G systems can provide seamless localization in various environments. However, significant challenges remain, including network interoperability, high overhead, Doppler effects, synchronization issues, and the need to mitigate overlapping sources of errors and hardware imperfections. Addressing these challenges requires the development of both innovative signal processing and networking techniques, along with aligned standardization and industrial solutions to ensure effective \ac{TN}-\ac{NTN} integration. Moving forward, the community should carefully study and decide on the optimal combination of integrated 6G enablers, the frequency of operation, and the level of integration for each application scenario to balance performance gains with solution complexity and operational constraints.

\bibliography{references}

\begin{thebibliography}{10}
\providecommand{\url}[1]{#1}
\csname url@samestyle\endcsname
\providecommand{\newblock}{\relax}
\providecommand{\bibinfo}[2]{#2}
\providecommand{\BIBentrySTDinterwordspacing}{\spaceskip=0pt\relax}
\providecommand{\BIBentryALTinterwordstretchfactor}{4}
\providecommand{\BIBentryALTinterwordspacing}{\spaceskip=\fontdimen2\font plus
\BIBentryALTinterwordstretchfactor\fontdimen3\font minus \fontdimen4\font\relax}
\providecommand{\BIBforeignlanguage}[2]{{%
\expandafter\ifx\csname l@#1\endcsname\relax
\typeout{** WARNING: IEEEtran.bst: No hyphenation pattern has been}%
\typeout{** loaded for the language `#1'. Using the pattern for}%
\typeout{** the default language instead.}%
\else
\language=\csname l@#1\endcsname
\fi
#2}}
\providecommand{\BIBdecl}{\relax}
\BIBdecl

\bibitem{dang2020should}
S.~Dang, O.~Amin, B.~Shihada, and M.-S. Alouini, ``What should {6G} be?'' \emph{Nat. Electron.}, vol.~3, no.~1, pp. 20--29, 2020.

\bibitem{9976205}
A.~Behravan, V.~Yajnanarayana, M.~F. Keskin, H.~Chen \emph{et~al.}, ``Positioning and sensing in {6G}: Gaps, challenges, and opportunities,'' \emph{IEEE Vehicular Technology Magazine}, vol.~18, no.~1, pp. 40--48, 2023.

\bibitem{9617565}
G.~Araniti, A.~Iera, S.~Pizzi, and F.~Rinaldi, ``Toward {6G} non-terrestrial networks,'' \emph{IEEE Network}, vol.~36, no.~1, pp. 113--120, 2022.

\bibitem{thz_6G_2019}
T.~S. Rappaport, Y.~Xing, O.~Kanhere, S.~Ju \emph{et~al.}, ``Wireless communications and applications above 100 {GHz}: Opportunities and challenges for {6G} and beyond,'' \emph{IEEE Access}, vol.~7, pp. 78\,729--78\,757, 2019.

\bibitem{6g_loc_sens_2024}
H.~Chen, M.~F. Keskin, A.~Sakhnini, N.~Decarli \emph{et~al.}, ``{6G} localization and sensing in the near field: Features, opportunities, and challenges,'' \emph{IEEE Wireless Communications}, vol.~31, no.~4, pp. 260--267, 2024.

\bibitem{guo2024integrated}
H.~Guo, H.~Wymeersch, B.~Makki, H.~Chen \emph{et~al.}, ``Integrated communication, localization, and sensing in {6G D-MIMO} networks,'' \emph{IEEE Wireless Communications}, 2024.

\bibitem{chen2023riss}
H.~Chen, H.~Kim, M.~Ammous, G.~Seco-Granados \emph{et~al.}, ``{RISs} and sidelink communications in smart cities: The key to seamless localization and sensing,'' \emph{IEEE Communications Magazine}, vol.~61, no.~8, pp. 140--146, 2023.

\bibitem{liang2024toward}
T.~Liang, T.~Zhang, and Q.~Zhang, ``Toward seamless localization and communication: {A} satellite-{UAV NTN} architecture,'' \emph{IEEE Network}, 2024.

\bibitem{9193893}
F.~Rinaldi, H.-L. Maattanen, J.~Torsner, S.~Pizzi \emph{et~al.}, ``Non-terrestrial networks in {5G} \& beyond: A survey,'' \emph{IEEE Access}, vol.~8, pp. 165\,178--165\,200, 2020.

\bibitem{1G_5G_survey_2018}
J.~A. Del Peral-Rosado, R.~Raulefs, J.~A. Lopez-Salcedo, and G.~Seco-Granados, ``\BIBforeignlanguage{en}{Survey of {cellular} {mobile} {radio} {localization} {methods}: {From} {1G} to {5G}},'' \emph{\BIBforeignlanguage{en}{IEEE Communications Surveys \& Tutorials}}, vol.~20, no.~2, pp. 1124--1148, 2018.

\bibitem{wang_recent_2023}
Y.~Wang, S.~Huang, Y.~Yu, C.~Li \emph{et~al.}, ``Recent progress on {3GPP} {5G} {positioning},'' in \emph{2023 {IEEE} 97th {Vehicular} {Technology} {Conference} ({VTC2023}-{Spring})}.\hskip 1em plus 0.5em minus 0.4em\relax Florence, Italy: IEEE, Jun. 2023, pp. 1--6.

\bibitem{10355106}
H.~K. Dureppagari, C.~Saha, H.~S. Dhillon, and R.~M. Buehrer, ``{NTN}-based {6G} localization: Vision, role of {LEOs}, and open problems,'' \emph{IEEE Wireless Commun.}, vol.~30, no.~6, pp. 44--51, 2023.

\bibitem{10714965}
F.~Fabra, D.~Egea–Roca, J.~A. López–Salcedo, and G.~Seco–Granados, ``Analysis on signals for {LEO-PNT} beyond {GNSS},'' in \emph{2024 32nd European Signal Processing Conference (EUSIPCO)}, 2024, pp. 1237--1241.

\bibitem{10542356}
W.~Stock, R.~T. Schwarz, C.~A. Hofmann, and A.~Knopp, ``Survey on opportunistic {PNT} with signals from {LEO} communication satellites,'' \emph{IEEE Communications Surveys \& Tutorials}, pp. 1--1, 2024.

\bibitem{Zheng2024LEO}
P.~Zheng, X.~Liu, and T.~Y. Al-Naffouri, ``{LEO}- and {RIS}-empowered user tracking: A {Riemannian} manifold approach,'' \emph{IEEE J. Sel. Areas Commun.}, vol.~42, no.~12, pp. 3445--3461, 2024.

\end{thebibliography}
\bibliographystyle{IEEEtran}
\end{document}